# The Paris Meudon ground based support to the NASA Solar Maximum Mission in the eighties

*Jean-Marie Malherbe*
Observatoire de Paris, PSL Research University, LIRA, France
Jean-Marie.Malherbe@obspm.fr; ORCID: https://orcid.org/0000-0002-4180-3729
Pierre.Mein@obspm.fr
21 May 2025

**ABSTRACT**

The Solar Maximum Mission of NASA was one of the first satellites with on board digitization of observations. It was launched for the solar maximum of cycle 21 (1980) in order to study the solar activity. It carried many instruments, such as coronagraphs, X and γ ray detectors, an Ultra Violet spectrometer and a radiometer. Ground based support was offered by many institutes, such as Paris Meudon observatory under the form of systematic observations or coordinated campaigns with specific instruments. We present here the Meudon Solar Tower (MST) and magnetograph which offered in the eighties a major contribution with observations of velocity and magnetic fields of the photosphere and chromosphere, while SMM was observing the transition region and corona above.

**INTRODUCTION**

The photosphere and chromosphere constitute the "cold" atmosphere (6000-8000 K) of the Sun with absorption lines formed in the visible part of the spectrum, which were historically observed with ground based instruments. The space era, in the sixties, open new windows on the Universe with access to UV and X-γ radiation, making the hot and outer regions of the Sun observable in emission lines of ionized atoms (the corona at $10^6$ K and the thin chromosphere-corona transition layer at $10^5$ K). In coronal physics, we can notice the growing competition from space, at first instruments using 35 mm films, such as the white light coronagraph onboard SKYLAB/NASA (Apollo Telescope Mount platform) from 1973 to 1979 (MacQueen *et al*, 1974), then those equipped with vidicon tubes, such as the SMM coronagraph, with real time digitization (MacQueen *et al*, 1980). NASA's Solar Maximum Mission satellite only operated for six months in 1980 (March-September) and unfortunately failed. It was successfully repaired in April 1984 by a spectacular extravehicular spacewalk from astronauts of the American space shuttle, and operated until 1989 with seven instruments (Vial, 1984). X-ray and γ-ray spectrometers allowed to study solar flares; coronal mass ejections were studied with the coronagraph, in collaboration with those of the Pic du Midi, the Nançay radioheliograph (169-470 MHz) and the monochromatic full disk Hα heliograph in Meudon. All aspects of solar activity (active regions, filaments, eruptions) were investigated by the Ultra Violet Spectrometer and Polarimeter (UVSP), which was particularly supported by Meudon magnetograph and MST, presented in this paper.

**I - MEUDON SOLAR TOWER (MST)**

This instrument was completed in 1969. Figure 1 provides a cross section of the instrument showing details of the beam. The concave mirror (60 cm diameter, F = 45 m) is located at the bottom of the tower. The beam is folded by two flat mirrors and forms a 42 cm image of the Sun at the telescope focus. There is a two mirror coelostat at the top of the tower (Figure 2). The number of flat mirrors, some with varying incidence, is not in favour of polarimetric measurements, Meudon magnetograph (section III) was more convenient for this purpose.

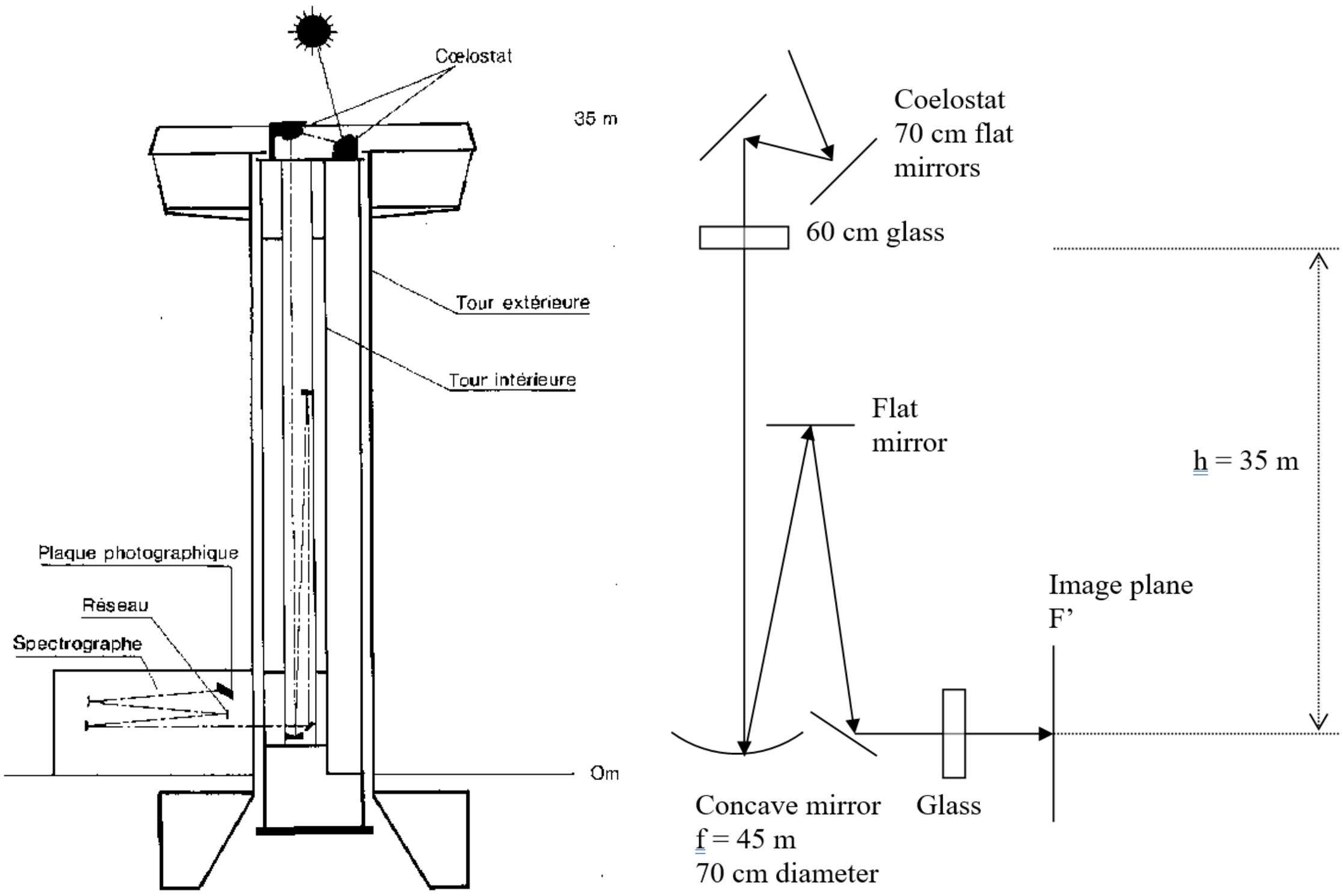

*Figure 1: The 60 cm / F = 45 m vertical telescope of Meudon Solar Tower is fed by a two mirror coelostat located at 35 m above the ground. Courtesy OP.*

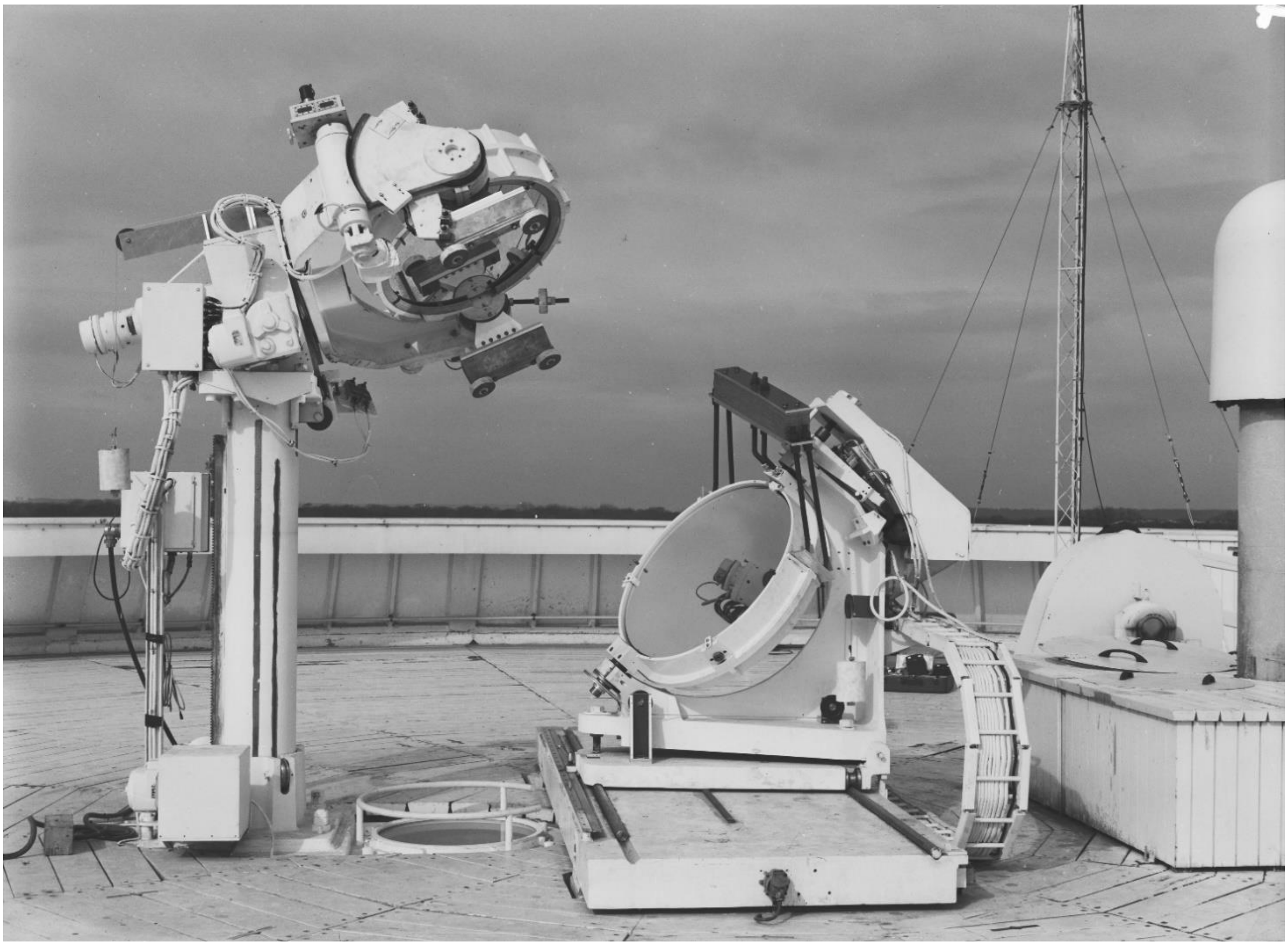

*Figure 2: The two mirror coelostat at the top of the tower. The telescope is vertical. Courtesy OP.*

The spectrograph (Figure 3) is made of a 300 grooves/mm grating (63°26’ blaze angle) and three concave mirrors of 14 m focal length: a collimator and two camera mirrors. The typical dispersion is 10 mm/Å but depends on wavelength. At the epoch of international campaigns with SMM, the detector was the photographic emulsion in the form of 70 mm films (45 m rolls, about 450 [6 x 9] $cm^2$ or 900 [6 x 4.5] $cm^2$ images recorded by Coleman cameras). The films were digitized with the Perkin Elmer PDS 1010 micro densitometer of Institut d’Optique (Orsay, France) in raster mode; it was connected to a Digital PDP 8 machine, pixels were written on 800 bpi magnetic tapes which were later processed by the VAX 11/780 in Meudon and archived on 1600/6250 bpi tapes.

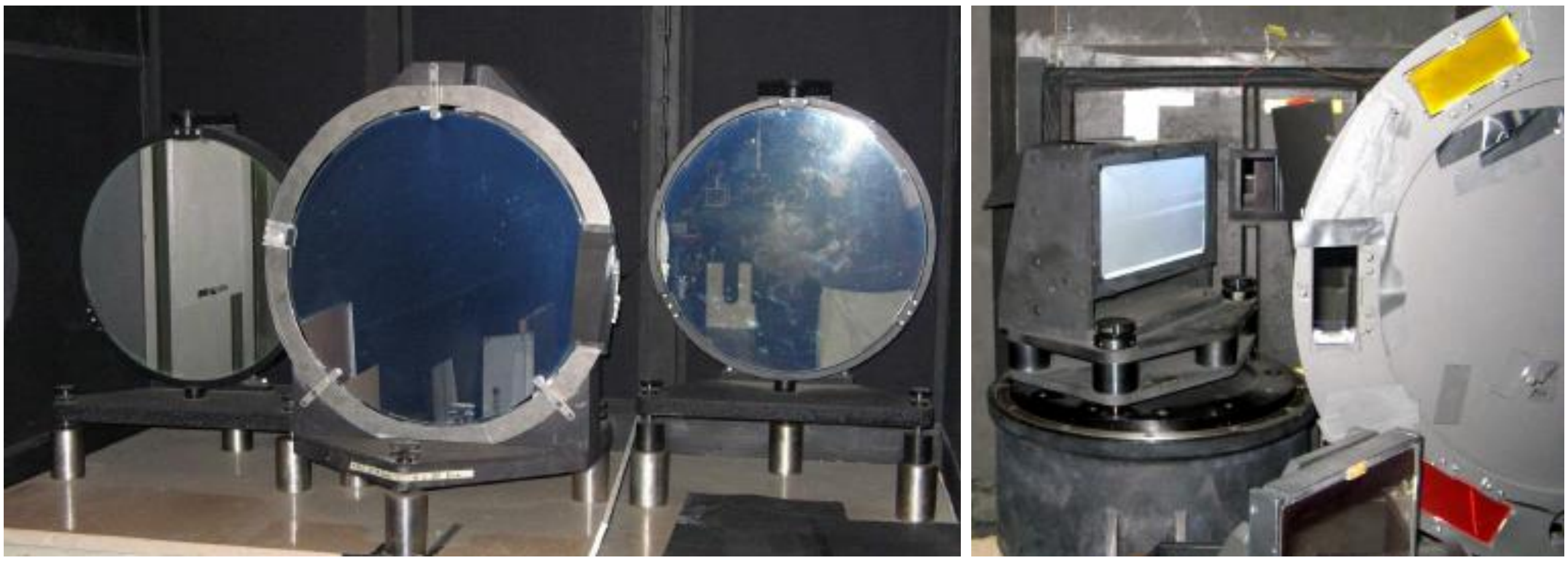

*Figure 3 : left : the collimator and two camera mirrors. Right : the grating and interference filters (100 Å FWHM) to select orders, from 5 (IR) to 15 (blue). Hα was at order 9. Courtesy OP.*

Campaigns were coordinated with the UVSP instrument onboard SMM, a spectrometer providing monochromatic images of the chromosphere corona transition region at $10^5$ K. The UVSP used four photomultipliers to observe several lines in raster mode, allowing to produce 2D spectroheliograms of regions in lines such as CIV; it had also the possibility to deliver dopplergrams with two photomultipliers observing simultaneously the red and blue wings of CIV, the intensity difference being proportional to the line of sight (LOS) velocity. For that reason, the spectrograph of MST was also able to work in fast 2D imaging spectroscopy mode, allowing to produce at high cadence spectroheliograms and dopplergrams of the chromosphere in Hα at 8000 K. This mode is the Multichannel Subtractive Double Pass (MSDP) explained in Section II.

## II - THE MSDP SPECTROGRAPH AT MST

The Multichannel Subtractive Double Pass (MSDP) is an imaging spectroscopy technique introduced by Mein (1977 and Figure 4). It is based on a slicer which provides line profiles with N sampling points (or N channels) over a 2D field of view (FOV); for that purpose, the MSDP uses a rectangular entrance window instead of a thin slit. The technique was progressively developed and implemented on many telescopes (Mein *et al*, 2021). The first instrument (N = 9 channels) was incorporated to the 14 m spectrograph of MST. It was mainly running with the Hα line in order to study the dynamics (via the Doppler effect) of chromospheric features (filaments, prominences, active regions, flares). The core of the MSDP is the slicer (Figure 5), a beam splitter-shifter using an entrance multi-slit to select N channels in the spectrum plus deviation prisms to reinject the N beams into the spectrograph for a second subtractive pass.

At the output of the spectrograph (Figure 6), the MSDP forms a spectra image composed of N channels of the same field of view (FOV); the wavelength function $\lambda_n(x)$ of each channel n is provided by Figure 7. If x is the abscissa along the entrance window, $\lambda_n(x) = \lambda_0 + x/d + n\,\Delta\lambda$, where d is the spectrograph dispersion (mm/Å) and $\Delta\lambda = s/d$ is the wavelength step between two consecutive channels, s being the slit step of the multi-slit slicer.

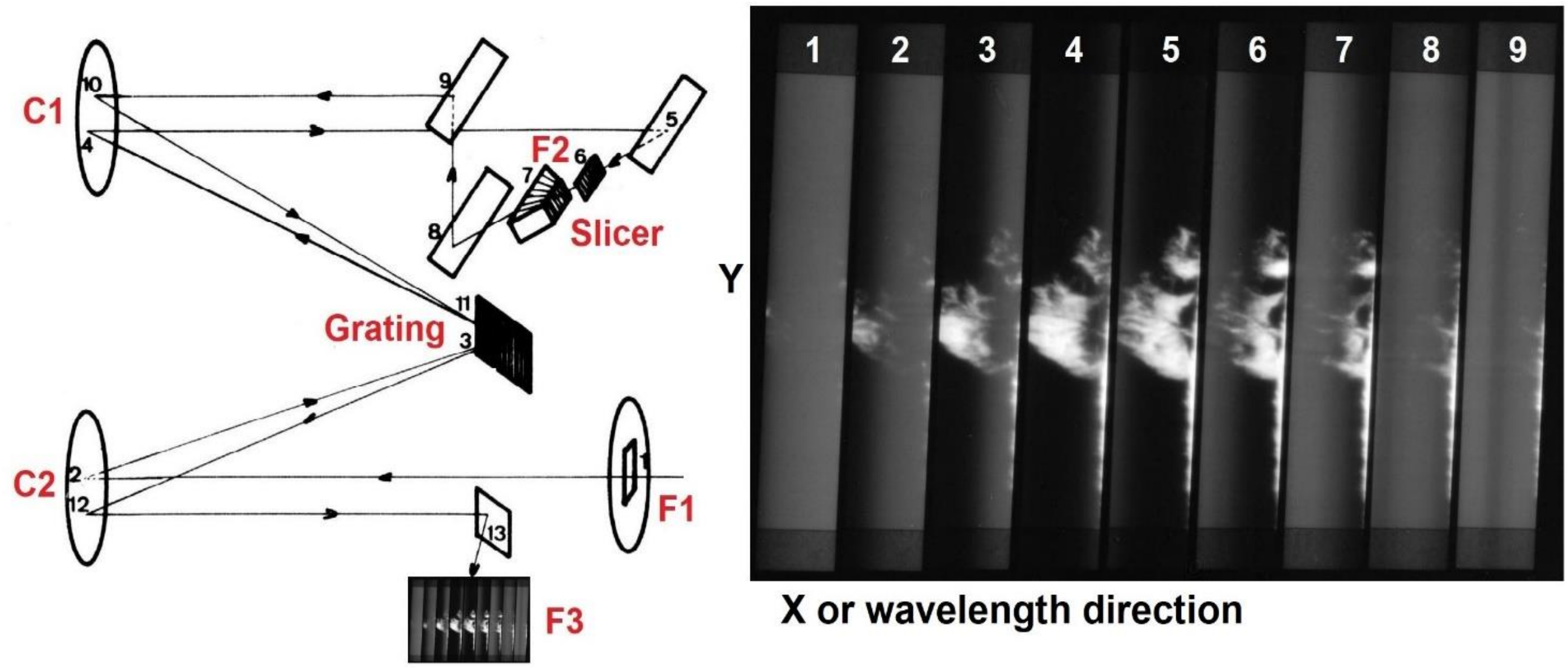

*Figure 4: the MSDP technique uses a classical spectrograph. F1 is the rectangular entrance window. C1 and C2 are the collimator and camera mirror. F2 is the spectrum, with the slicer (beam splitter shifter). The MSDP is characterized by a second subtractive second pass on the grating. F3 is the final spectra image with N rectangular channels. At right, an Hα spectra image of a prominence with N = 9 channels of the same field of view. Courtesy OP.*

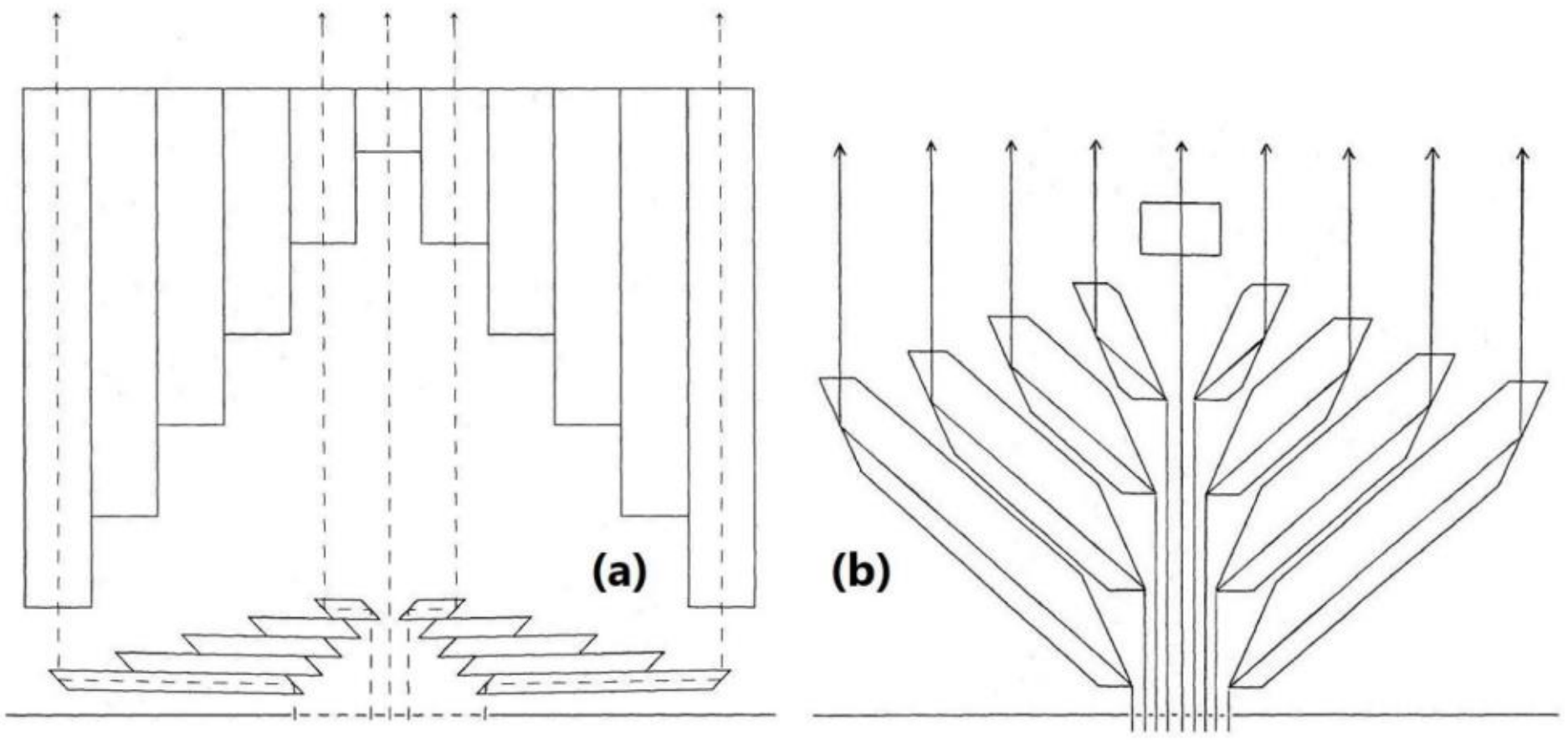

*Figure 5: the MSDP slicers a or b are based on multi slits in the spectrum and prism beam shifters. Courtesy OP.*

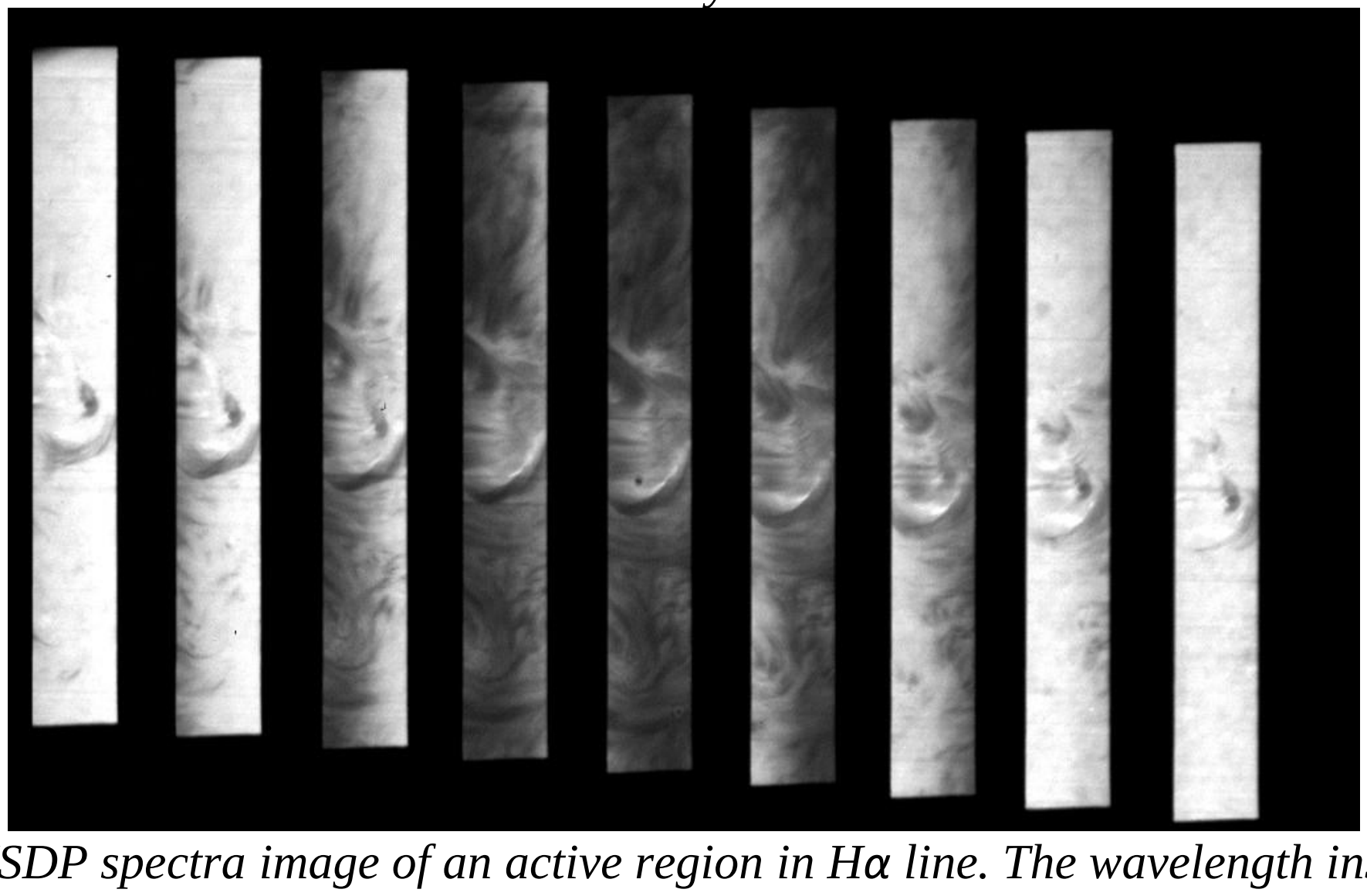

*Figure 6: MSDP spectra image of an active region in Hα line. The wavelength inside channel number n at abscissa x is equal to $\lambda(x) = \lambda_0 + x/d + n\,\Delta\lambda$, where d is the spectrograph dispersion (mm/Å) and $\Delta\lambda = s/d$ is the wavelength step between two consecutive channels, s being the slit width of the slicer. Courtesy OP.*

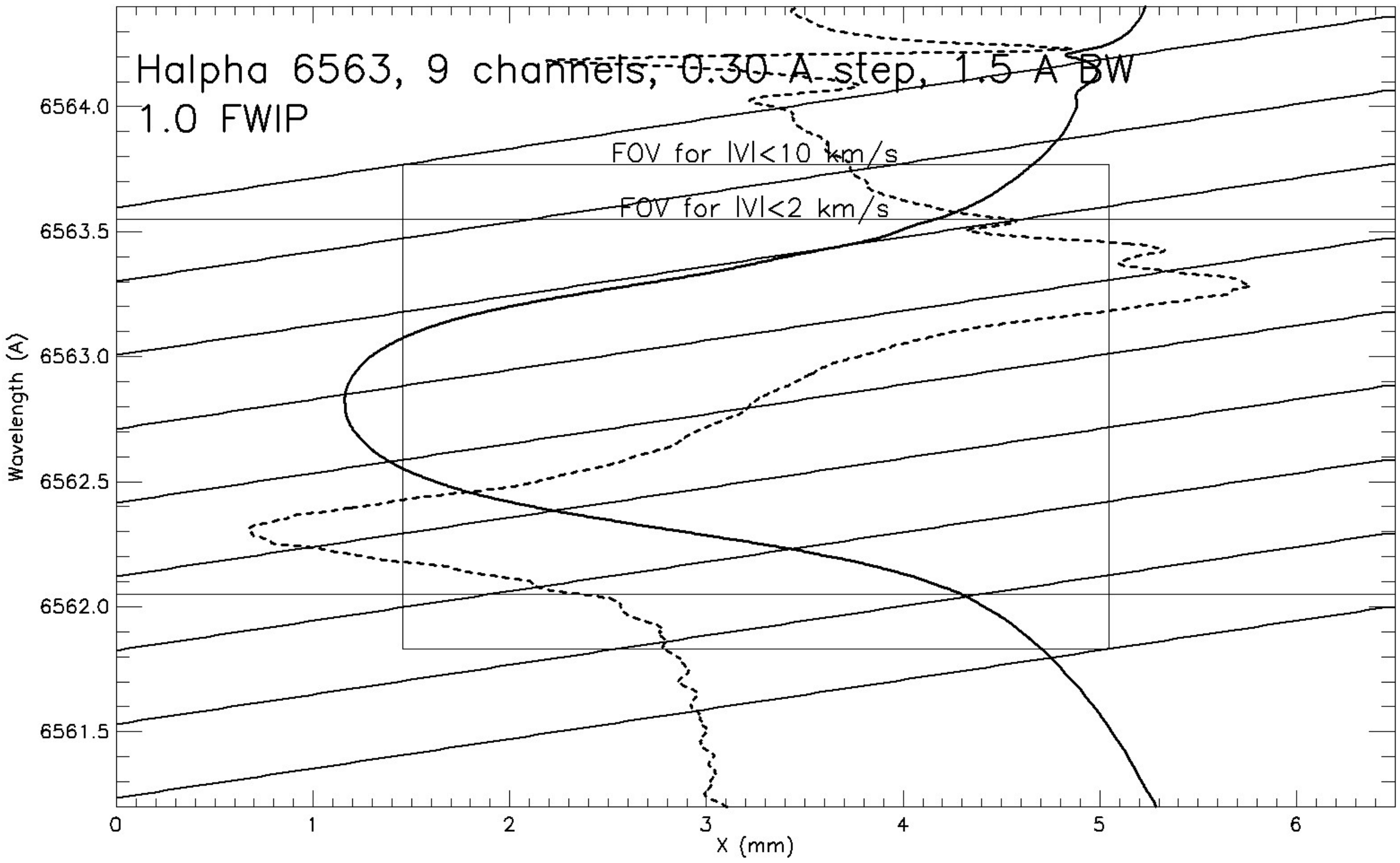

*Figure 7: The wavelength for Hα line inside channel number n at abscissa x is equal to $\lambda(x) = \lambda_0 + x/d + n\,\Delta\lambda$, where d = 8 mm/Å is the spectrograph dispersion and $\Delta\lambda = s/d = 0.3$ Å is the wavelength step between two consecutive channels, s = 2.5 mm being the step of the slicer. The solid line is the atlas profile of Hα at disk centre. The dashed line is the first derivative. Inflexion points are distant of 1.0 Å (FWIP). We imposed the bandwidth BW = 1.5 Å. With this condition, LOS velocities below 2 km/s are measurable everywhere in the full FOV (0 < x < 6.5 mm); but LOS velocities up to 10 km/s require more bandwidth (BW = 1.9 Å), and consequently the FOV is reduced (rectangle 1.4 mm < x < 5.1 mm). Courtesy OP.*

Figures 8 & 9 illustrate typical mappings of the MSDP in Hα line. After 1980, it was possible to produce pseudo coloured maps using the image processor COMTAL vision one/20 (512 x 512 pixels) which was connected to the Unibus of the VAX 11/780.

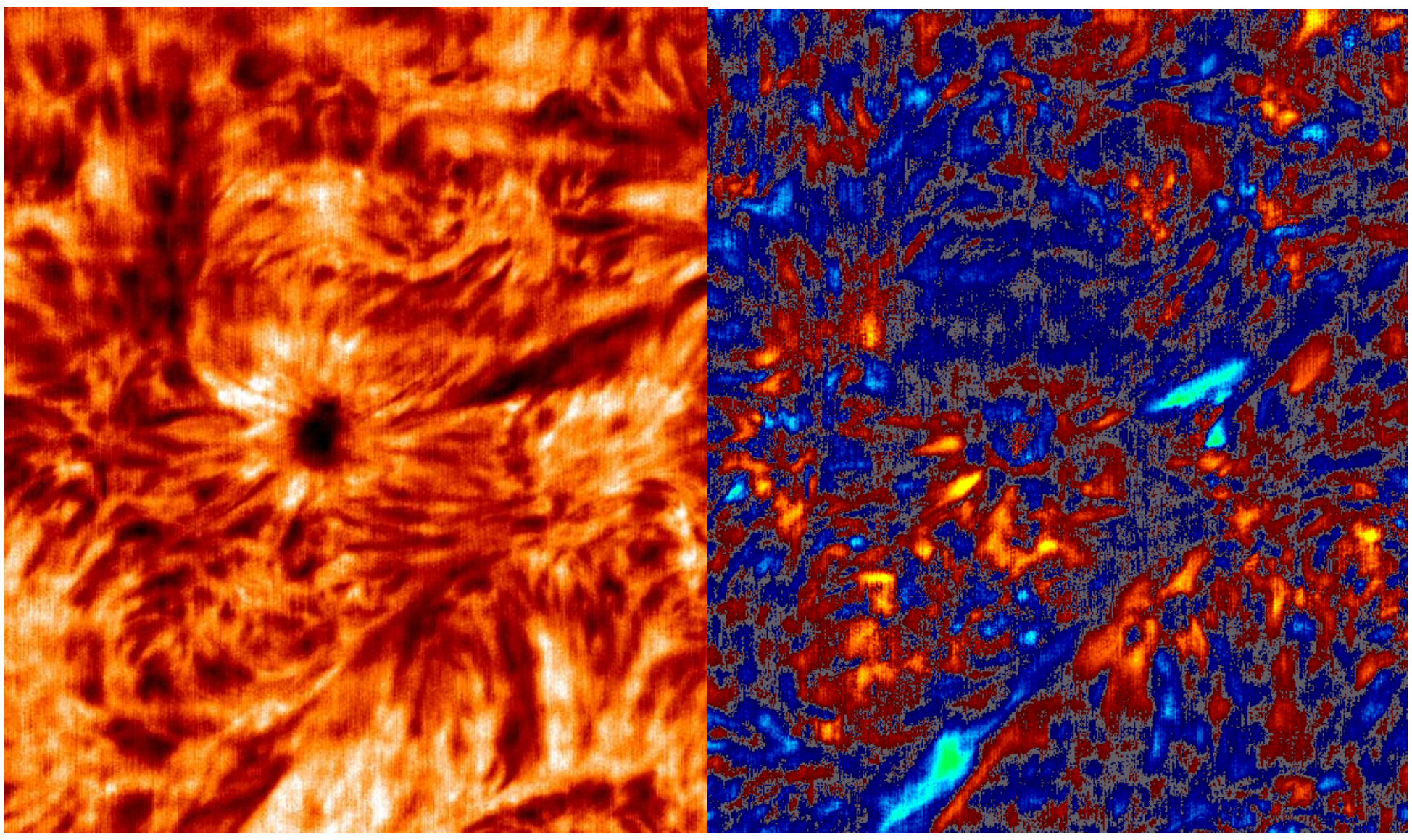

*Figure 8: Active region in Hα line (left: intensity fluctuations, right: dopplershifts measured at ± 0.3 Å from the line core, below the inflexion points, in the chromosphere). Courtesy OP.*

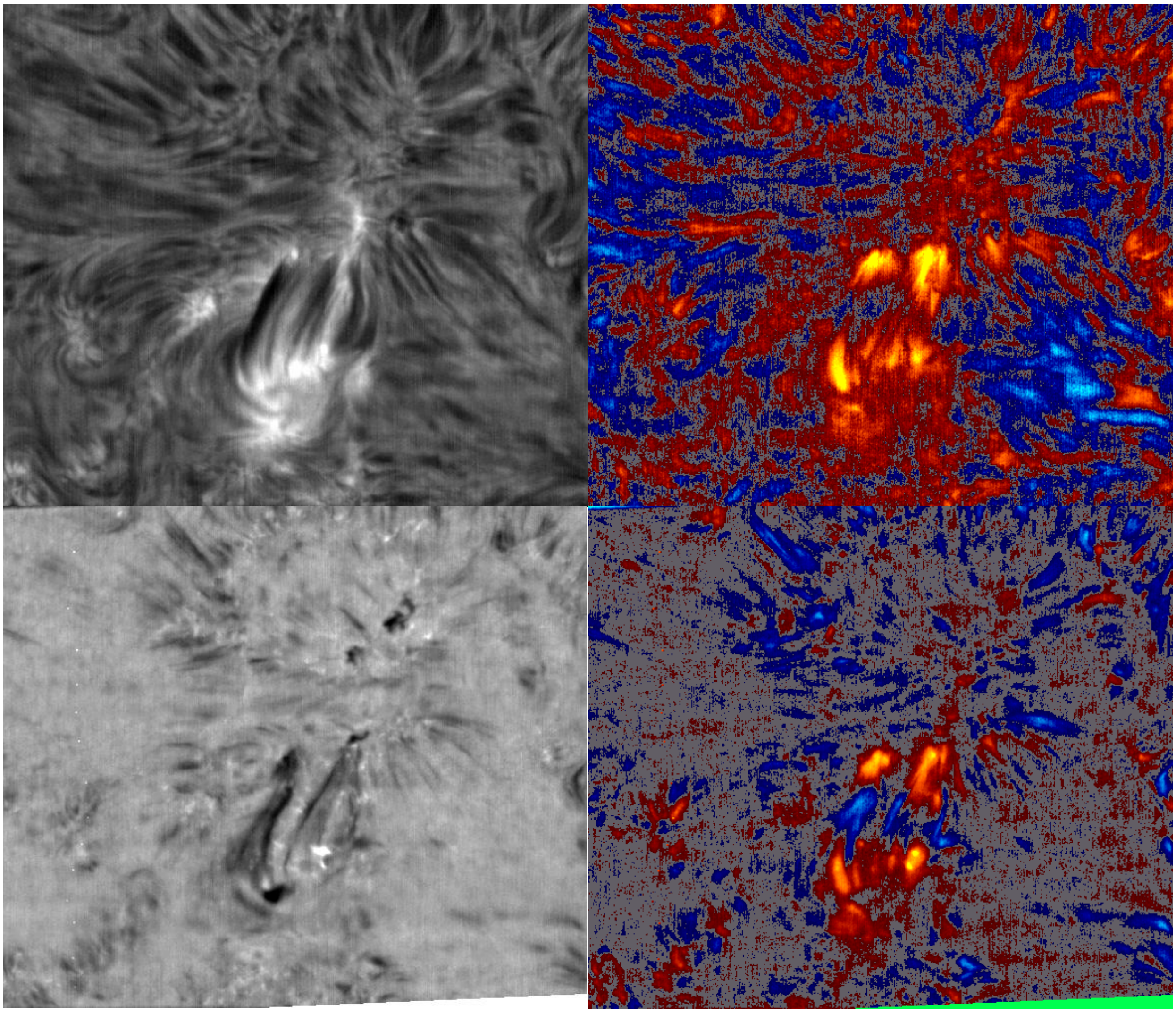

*Figure 9: Arch filament system in Hα line (left: intensity fluctuations, right: dopplershifts). Intensities and LOS velocities were measured at two altitudes using the bisector method at ± 0.3 Å from the line core, below the inflexion points, in the chromosphere (top) and at ± 0.6 Å from the line core, above the inflexion points, in the high photosphere (bottom). Courtesy OP.*

## III - MEUDON MAGNETOGRAPH

The measurements of magnetic fields, through the Zeeman effect, are strongly associated to the analysis of the spectral line polarization, such as FeI lines for the photosphere, NaI or MgI lines for the low chromosphere or CaII lines for the chromosphere. Meudon had a long experience in polarimetric techniques, the development of which started with Lyot (the monochromatic birefringent filter, 1944) and continued with many astronomers such as Dollfus, Charvin, Semel or Rayrole.

The LOS magnetic field B// can be obtained from the circular polarization; the circular polarization rate V/I is proportional to the ratio ($\Delta\lambda_{B//}$ / Δλ). $\Delta\lambda_{B//}$ = [e / (4π m C)] B// g* $\lambda^2$ is the Zeeman splitting and Δλ the width of the spectral line. g* is the equivalent Lande factor. The transverse magnetic field B_|_ is much more difficult to measure, because the linear polarization rate is proportional to the ratio $(\Delta\lambda_{B_|_} / \Delta\lambda)^2$, it is a second order, with $\Delta\lambda_{B_|_}$ = [e / (4π m C)] B_|_ g* $\lambda^2$. The determination of the linear polarization is strongly affected by instrumental parasitic polarization induced by reflections on mirrors, this is the reason why most measurements in the second half of the twentieth century were done in circular polarization, since many solar telescopes were fed by 2-mirror coelostats. At Meudon, the magnetography started in the sixties with a 40 cm, 7 m focal length Newtonian telescope fed by a siderostat (Figure 10). Only circular polarization analysis was done.

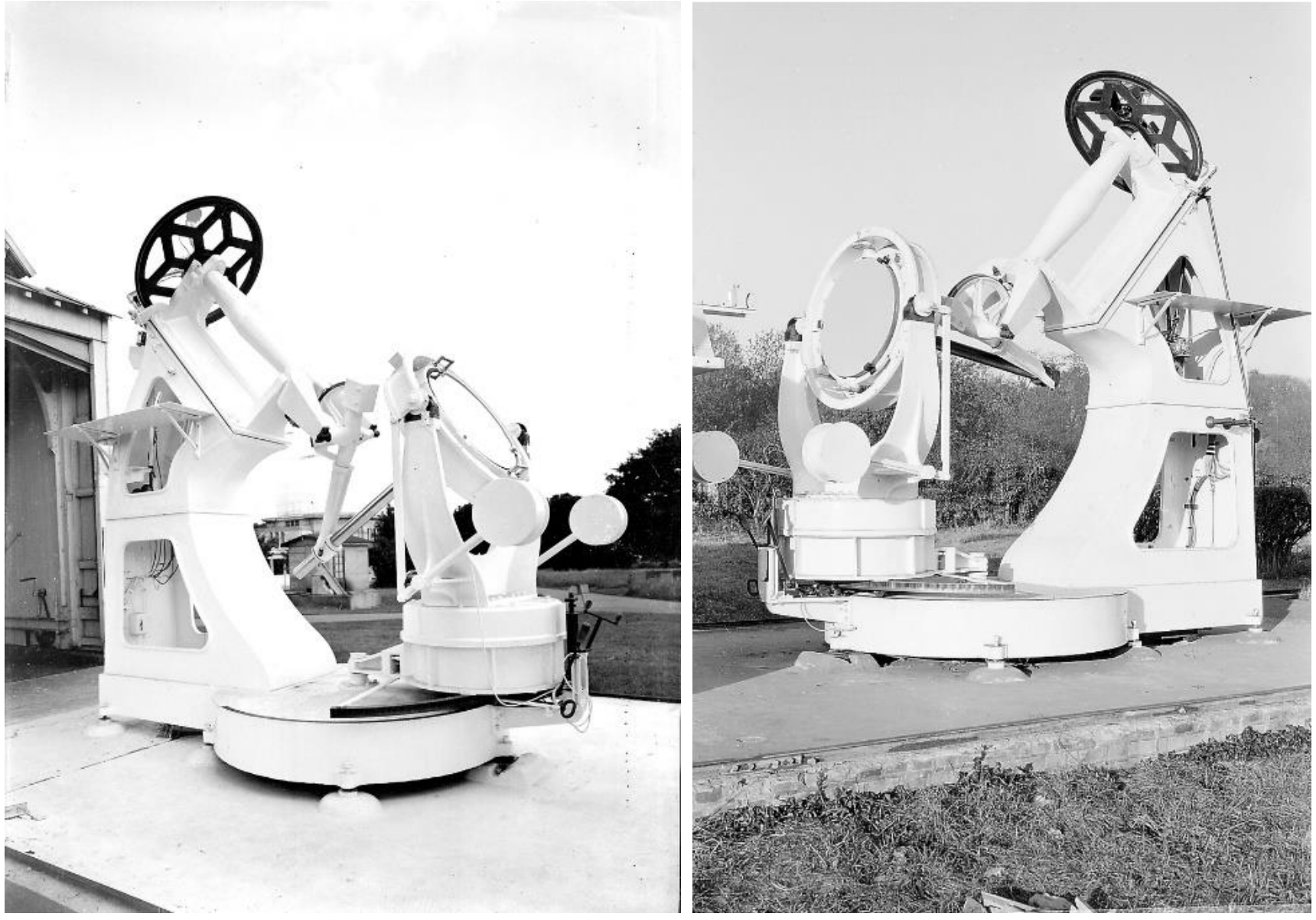

*Figure 10: the 85 cm Foucault siderostat feeding the Meudon magnetograph. Courtesy OP.*

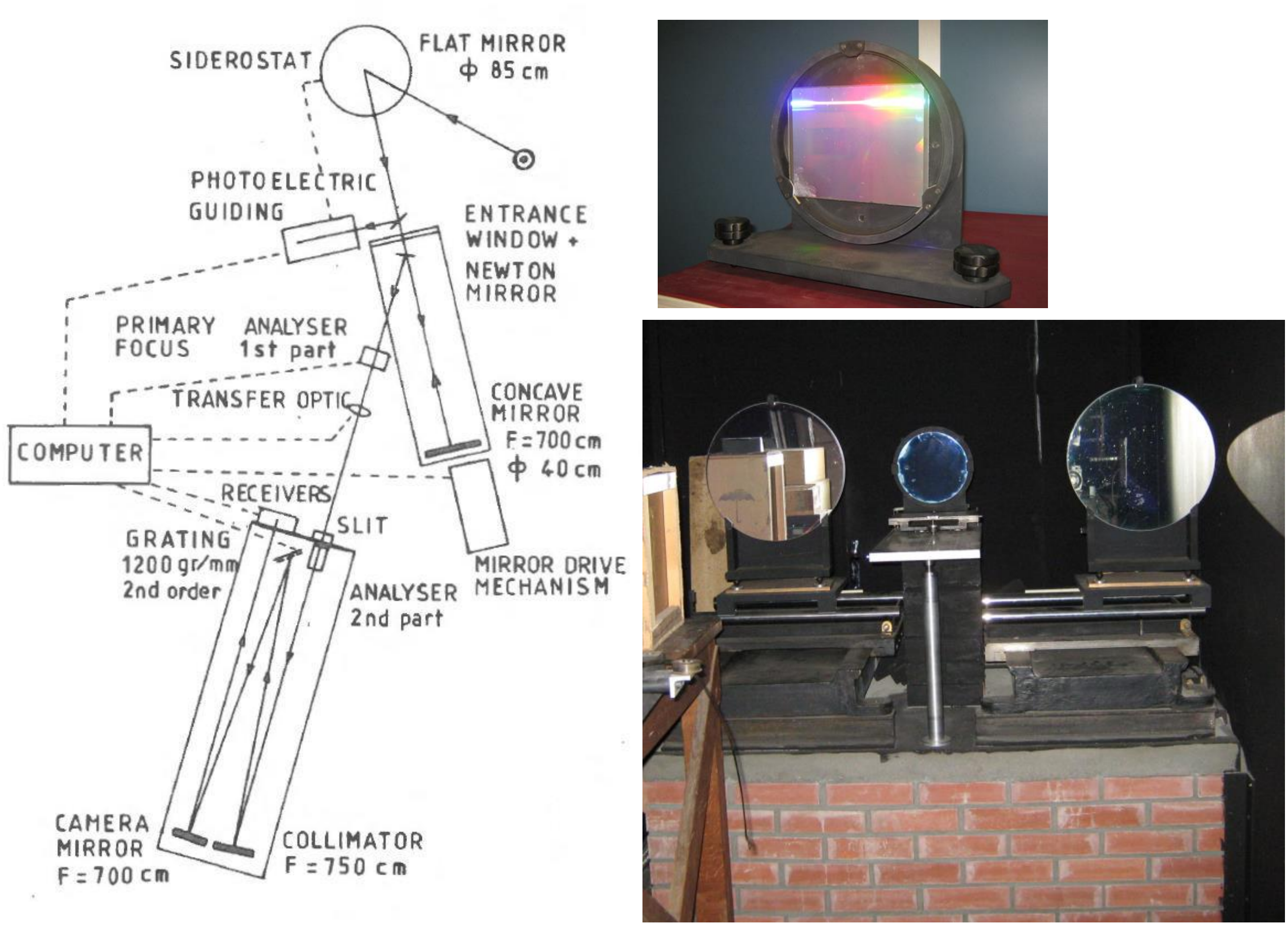

*Figure 11: left: the telescope and spectrograph of Meudon magnetograph in 1980 (after Rayrole, 1980). Right: the collimator (f = 7.5 m) and two camera mirrors (f = 7 m). Top: the 1200 grooves/mm grating working in the second order. Courtesy OP.*

Measurements of magnetic fields began in 1965 using a Hale Nicholson grid located in front of the spectrograph slit and composed of 3" alternate strips of quarter and three quarter wave material (Michard & Rayrole, 1965). But this single beam approach had the disadvantage to provide I+V and I-V spectra which were not cospatial. Rayrole (1967) introduced a more sophisticated polarimeter (quarter waveplate and birefringent calcite beam splitter shifter) providing two cospatial beams. In both versions, the spectra were recorded on photographic plates.

Figures 11 & 12 show the improved version of the spectrograph in 1980. One camera mirror was used for lines sensitive to the magnetic field (high equivalent Lande factor g*) while the second chamber was for non magnetic lines (g* = 0) such as FeI 5576 Å. Figure 13 details the output: the spectra of the Zeeman components (left and right circular polarization) were focused on two Reticon diode linear arrays (256 pixels each), orthogonal to the dispersion direction. The Reticon arrays were motorized in translation and moved to scan the lines (Figure 14). A micro computer Texas Instruments calculated in real time differential profiles to locate precisely in wavelength the Zeeman components, $\lambda_1$ and $\lambda_2$. The radial velocity is proportional to $(\lambda_1 + \lambda_2)/2$ (the Dopplershift) and the LOS magnetic field to $(\lambda_1 - \lambda_2)/2$ (the Zeeman splitting). Maps of magnetic fields were stored on 8 inches floppy disks (512 Ko), and then brought to the COMTAL image processor for false colour representation (Figure 15).

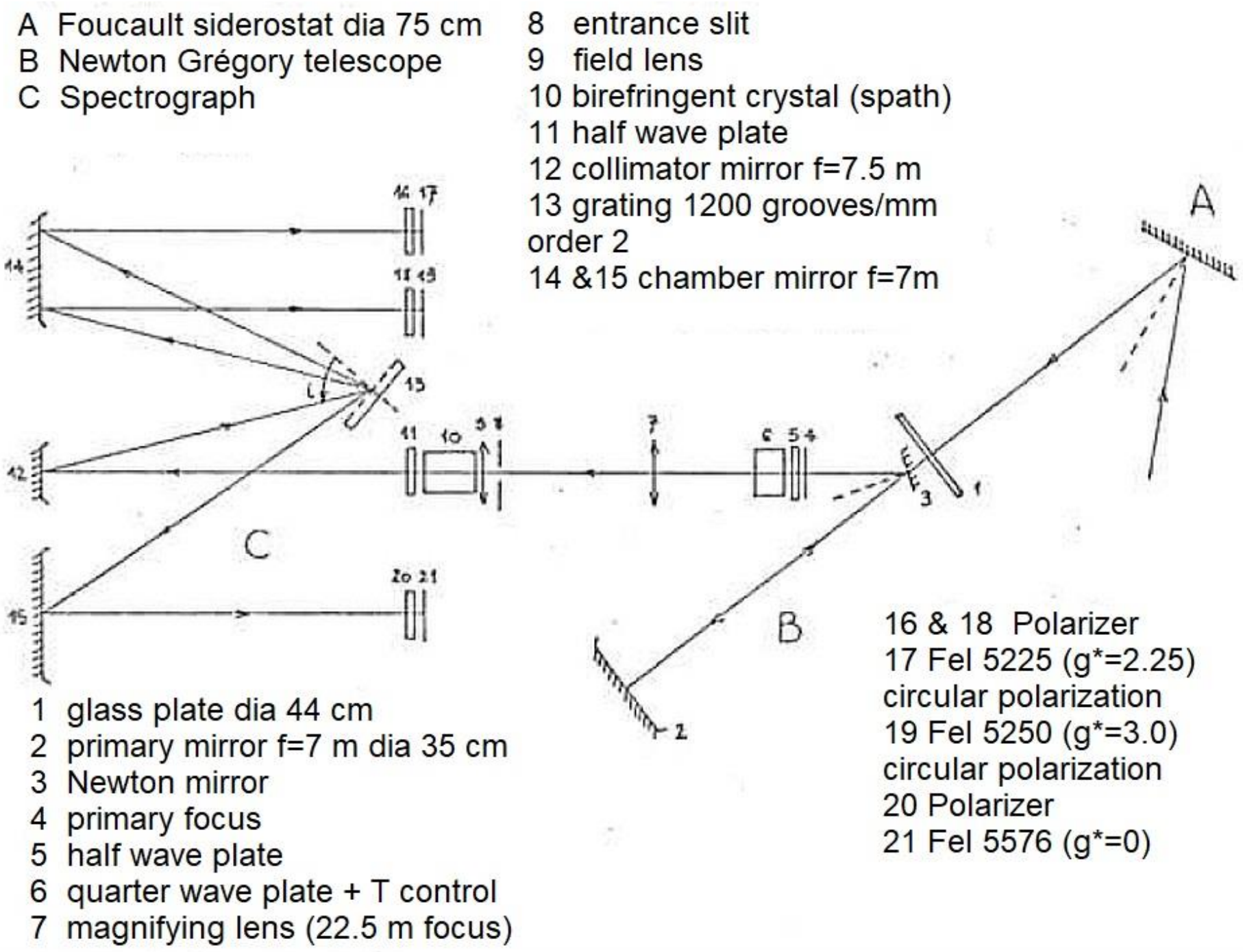

*Figure 12: the telescope and spectrograph of Meudon magnetograph in 1980. Courtesy OP.*

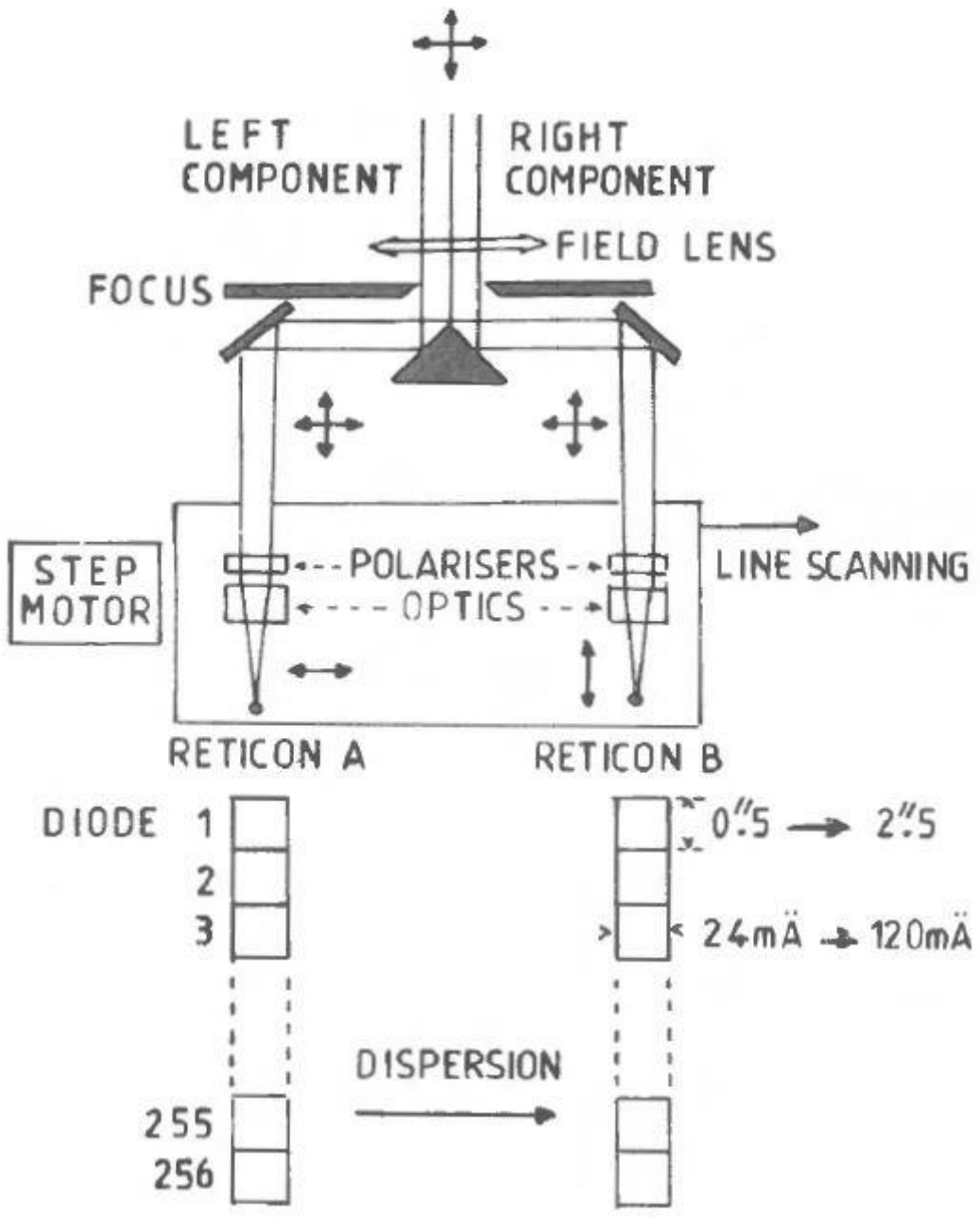

*Figure 13: the spectrograph exit and the two Reticon diode arrays to scan the Zeeman components. After Rayrole, 1980. Courtesy OP.*

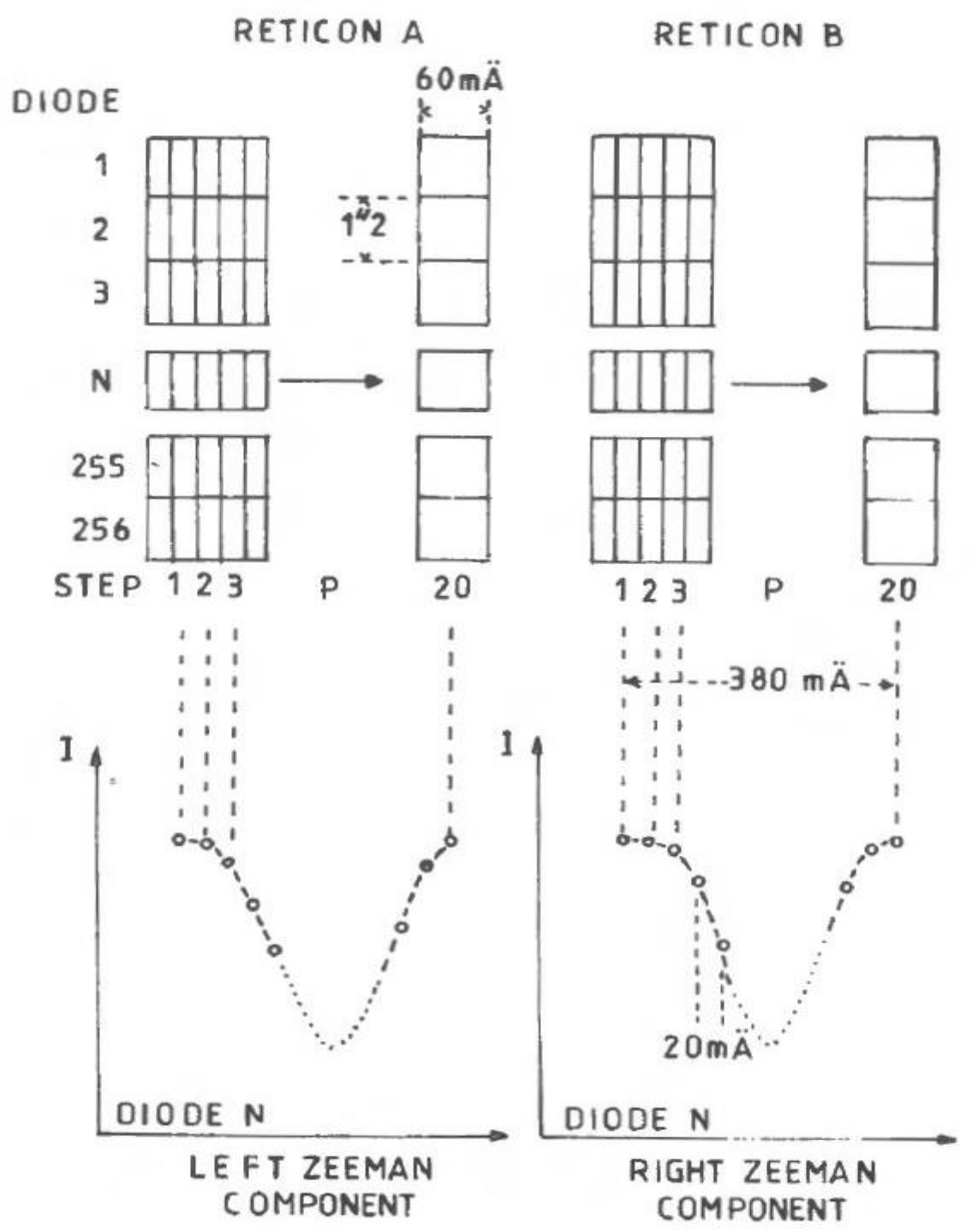

*Figure 14: the two Reticon diode arrays scan the left and right circular polarization spectra of the Zeeman components. After Rayrole, 1980, and courtesy OP.*

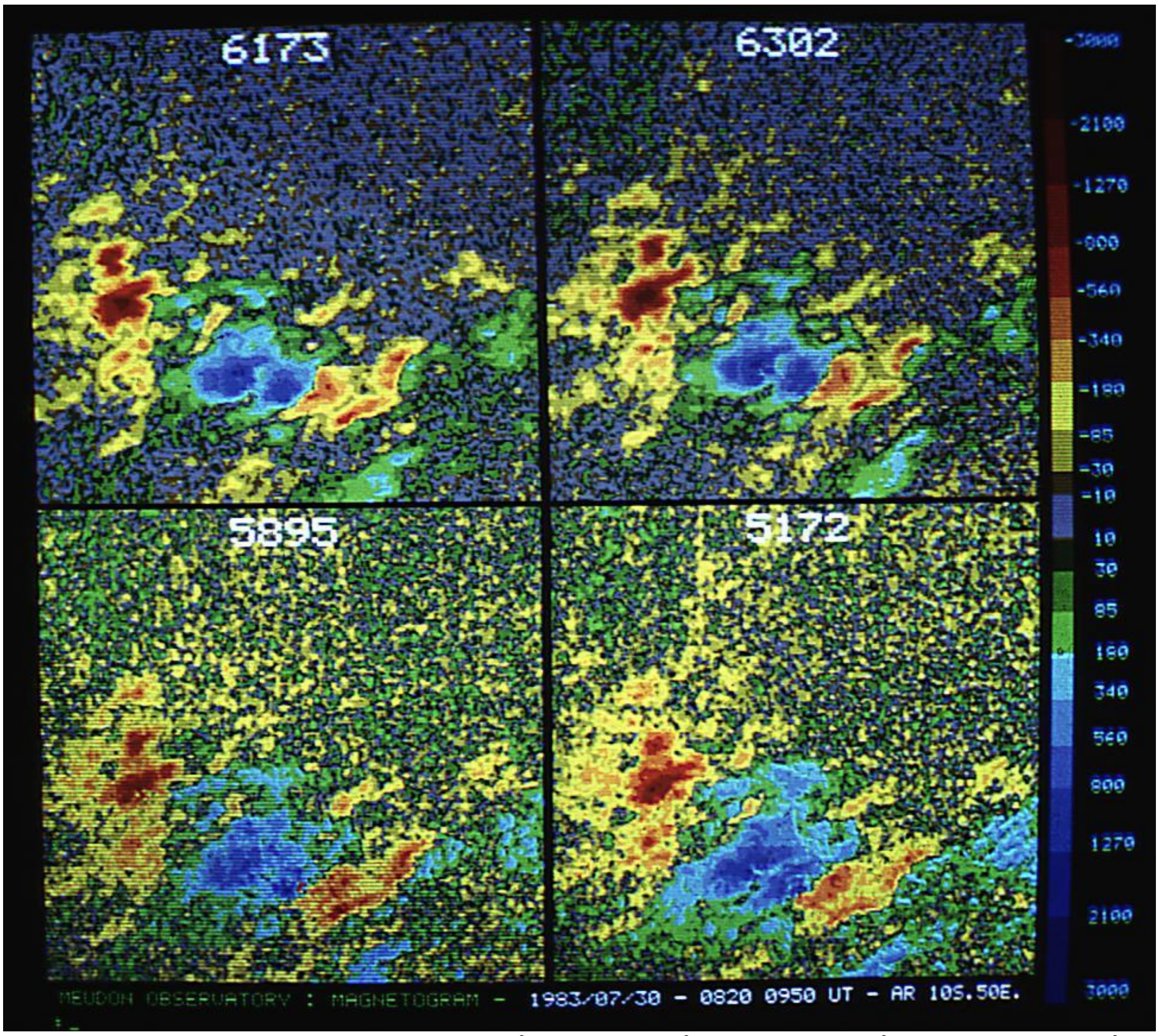

*Figure 15: magnetic field maps in FeI 6173 Å, FeI 6302 Å, NaD1 5896 Å, and MgI 5173 Å, 30 July 1983. Active region 10°S, 50°E. Polarities in yellow-red/green-blue. Courtesy OP.*

## IV – A TYPICAL EXAMPLE OF COORDINATED OBSERVATIONS WITH SMM

The standard observing mode of the LOS velocities with the UVSP/SMM consisted in measuring the red and blue wings of UV emission lines. As the precision was rather limited with only two samples in the line profile, Simon *et al* (1982) proposed an improved method (The Double Dopplergram Determination or DDD) with 4 points allowing to measure velocities up to 80 km/s instead of 30 km/s with the usual procedure.

Figure 16 displays coordinated observations with the UVSP/SMM in the CIV 1548 Å emission line (intensity and Dopplershift), the MSDP of MST in Hα (chromospheric absorption line, intensity and Dopplershift) and the Meudon magnetograph (LOS magnetic field of the photosphere). These observations (Schmieder *et al*, 1985) allowed us to study mass motions in an active region filament (8000 K) and in the transition zone around ($10^5$ K) in connection with the photospheric magnetic field below, showing the polarities and anchorage of magnetic field lines. Upward motions were evidenced in the filament (a dark and cold structure embedded in the low and hot corona) while downflows occurred at the footpoints (located at the ends of the filament in the chromosphere); this result was consistent with mass flowing along a flux rope and draining out the ends.

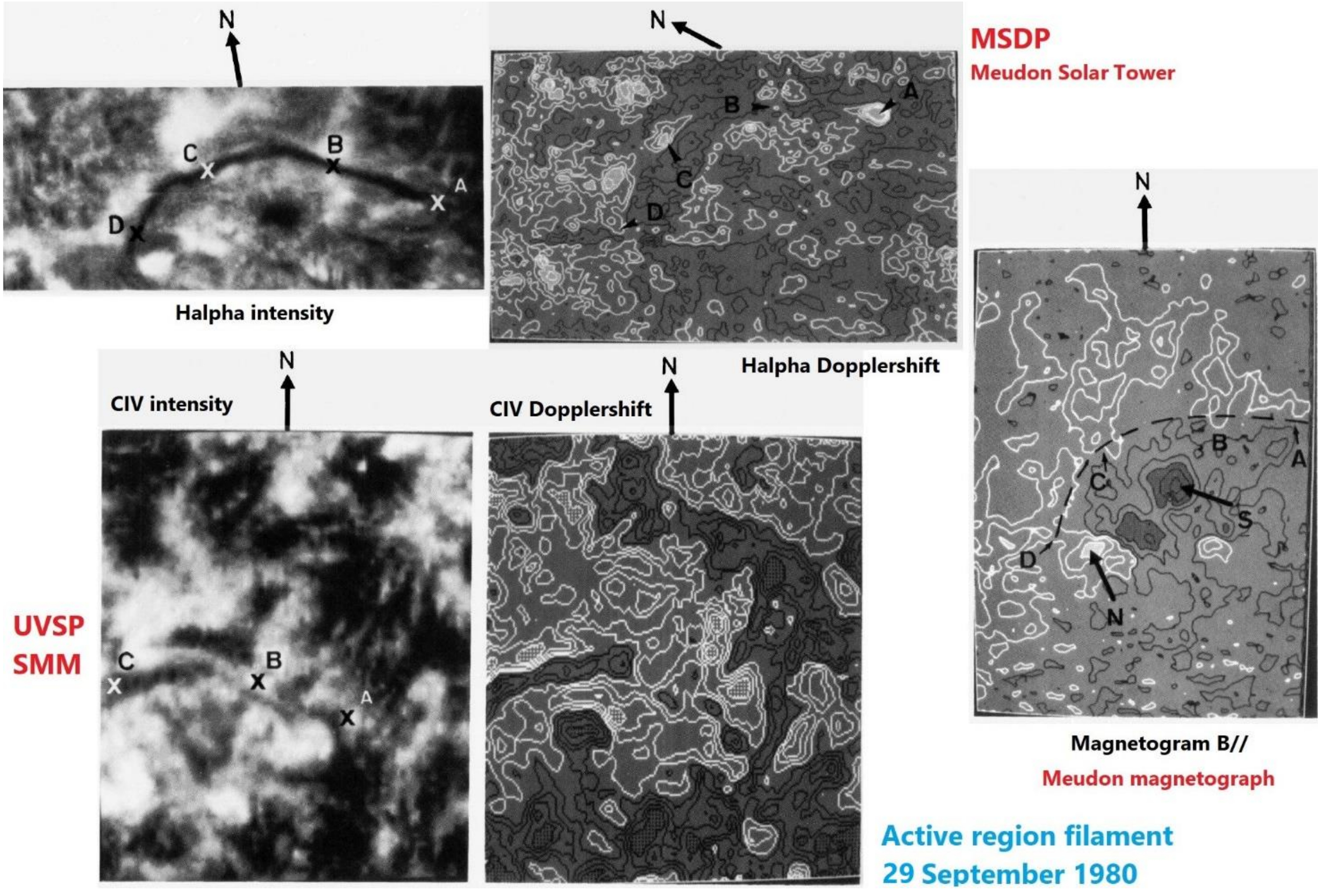

*Figure 16: coordinated observations with the UVSP onboard SMM in CIV line, the MSDP at Meudon Solar Tower in Hα and Meudon magnetograph. 29 September 1980. Courtesy OP.*

## CONCLUSION

Many coordinated campaigns were done in the eighties between SMM and ground base solar telescopes, allowing to link photospheric and chromospheric events to phenomena in higher and hotter layers above (chromposhere corona transition zone and corona), such as mass motions in active regions and filaments, flares and coronal ejections. SMM was at the source of a large number of space missions, such as Yohkoh (Solar A, JAXA, 1991), SOHO (ESA/NASA, 1996), Trace (NASA, 1998), Hinode (Solar B, JAXA/NASA, 2006), STEREO (NASA, 2006), SDO (NASA, 2010), IRIS (NASA, 2013) Solar Probe (NASA, 2018) and Solar Orbiter (ESA, 2020). American ground based facilities (such as Sacramento Peak, Kitt Peak) were followed by the giant 4 m DKIST (2020); THEMIS replaced Meudon instruments described here (1999); European astronomers are now preparing the renew of solar facilities in La Palma and Tenerife under the form of the 4 m EST telescope, a project for 2030 but not yet funded.

## MOVIES

In 1980, the MP4 format did not exist and it was necessary to use Super 8 or 16 mm film cameras to record the screen of image processors and produce movies of the solar activity. The two movies below come from digitized 16 mm films.

CIV1980.mp4
Shows typical events observed in CIV with the UVSP onboard SMM in 1980

Halpha1980.mp4
Shows typical events observed in Hα with the MSDP Meudon Solar Tower in 1980

## REFERENCES

MacQueen, R., Eddy, J., Gosling, J., Hildner, E., Munro, R., Newkirk, G., Poland, A., Ross, C., 1974, "the outer corona as observed from SkyLab: preliminary results", *ApJ*, 187, L85-L88

MacQueen, R., Csoeke-Poeckh A., Hildner, E., House, L., Reynolds, R., Stanger, A., Tepoel, H., Wagner, W., 1980, "the high altitude observatory coronagraph/polarimeter on the solar maximum mission", *Solar phys*., 65, 91-107

Mein P., "*Multi-channel subtractive spectrograph and filament observations*", 1977, Sol. Phys., 54, 45, https://ui.adsabs.harvard.edu/abs/1977SoPh...54...45M

Mein, P., Malherbe, J.M., Sayède, F., Rudawy, P., Phillips, K., Keenan, F., 2021, "*Four decades of advances from MSDP to S4I and SLED imaging spectrometers*", Solar Phys., 296, 30; see also arXiv:2101.03918, https://arxiv.org/abs/2101.03918

Michard, R., Rayrole, J., 1965, Observation systématique des champs magnétiques des centres d'activité à l'observatoire de Meudon, IAUS, 22, 169-172
Rayrole, J., 1967, Contribution à l'étude de la structure du champ magnétique dans les taches solaires, thesis, *An. Ap*., 30, 257-300

Rayrole, J., 1980, Magnetic field observations in Meudon observatory, *Japan France Seminar in Solar Physics*, 258-268

Schmieder, B., Malherbe, J.M., Poland, A., Simon, G., 1985, Dynamics of solar filaments – IV – structure and mass flow of an active region filament, *Astron. Astrophys*., 153, 64-70

Simon, G., Mein, P., Vial, J.C., Shine, R., Woodgate, B., 1982, Measurements of solar transition zone velocities and line broadening using the ultraviolet spectrometer and polarimeter on the solar maximum mission, , *Astron. Astrophys*., 115, 367-372